\documentstyle[prl,aps,epsf,floats]{revtex}

\begin{document}

\newcommand{\DR}{{\footnotesize{$\overline{{\rm DR}}$}} }
\newcommand{\MS}{{\footnotesize{$\overline{{\rm MS}}$}} }
\newcommand{\three}{{$SU(3) \times SU(2) \times U(1)$} }
\newcommand{\two}{{$SU(2) \times U(1)$} }
\newcommand{\be}{\begin{equation} }
\newcommand{\ee}{\end{equation} }
\newcommand{\bea}{\begin{eqnarray} }
\newcommand{\eea}{\end{eqnarray} }
\newcommand{\nn}{\nonumber}
\newcommand{\rref}{\rm ref}
\def\gsim{\raise.3ex\hbox{$\,>$\kern-.75em\lower1ex\hbox{$\sim$}\,}}
\def\lsim{\raise.3ex\hbox{$\,<$\kern-.75em\lower1ex\hbox{$\sim$}\,}}
\def\L{{\cal L}}
\def\O{{\cal O}}
\def\Tr{{\rm Tr\, }}

\twocolumn[
\hsize\textwidth\columnwidth\hsize
\csname@twocolumnfalse\endcsname
\title{Precision Observables and Electroweak Theories}
\author{Jonathan A.~Bagger, Adam F.~Falk and Morris Swartz\\[4pt]}
\address{\tighten{\it
Department of Physics and Astronomy,
The Johns Hopkins University,\\
3400 North Charles Street,
Baltimore, Maryland 21218 USA}\\[4pt] August, 1999}
\maketitle
\begin{abstract}
We compute the bounds from precision observables on
alternative theories of electroweak symmetry breaking.
We show that a cut-off as large as 3 TeV can be
accomodated by the present data, without unnatural
fine tuning.
\end{abstract}
\vspace{0.2in}
]\narrowtext

\section{Introduction}

During the past few years, precision measurements of
electroweak observables have probed the standard model
of particle physics to the 0.1\% level.  They now
give a 95\% C.L.\ upper bound of 230 GeV on the
mass of the standard model Higgs boson~\cite{morris}.
Precision measurements have also constrained many
alternative theories to the standard model.  For
example, they have ruled out many of the most naive
technicolor theories~\cite{peskin}.

The theory of effective Lagrangians provides a
convenient way to describe the low-energy effects
of new physics beyond the standard model.  One
approach is to take the standard model with a
fundamental Higgs boson and add a set of \three
invariant higher-dimensional operators, suppressed
by a scale $\Lambda$.  These operators are
assumed to be generated by new physics at the scale
$\Lambda$, beyond that of the usual standard model.
Because the effective theory includes a fundamental
Higgs boson, triviality gives the only upper bound
on the scale $\Lambda$.  This approach has recently
been used to study the Higgs mass limit that comes
from precision measurements. It was shown that the
new operators can raise the limit on the Higgs
mass as high as 400-500 GeV, barring unnatural
cancellations~\cite{chivukula}.

A second approach is to eliminate the Higgs entirely
and parametrize the present data in terms of the
standard model fields that have been discovered to
date.  In this approach, $\Lambda$ defines the scale
of the physics responsible for electroweak symmetry
breaking.  At low energies, all effects of this
physics can be described by a gauge invariant chiral
Lagrangian, in which the higher-dimensional operators
are suppressed by $\Lambda$.  This approach is valid
for energies $E \lsim \Lambda$.  General unitarity
considerations restrict $\Lambda \lsim 3$\,TeV.

In this letter we pursue this second approach
and focus on the physics of electroweak symmetry
breaking.  We will use the precision measurements
to constrain the coefficients of the leading higher
dimensional operators in the chiral Lagrangian, as
a function of the scale $\Lambda$.  We will find
that even for $\Lambda\approx3\,$TeV, the present
precision data can be accomodated without unnatural
fine tuning.

If $\Lambda \simeq 3$\,TeV, the physics of
electroweak symmetry breaking lies outside the
reach of the LEP and Tevatron colliders.  Our
analysis indicates that this possibility remains
open, despite the 230 GeV lower limit on the
mass of the standard model Higgs.  We shall
see that the data is perfectly consistent
with theories in which there are {\it no\/}
new particles below 3~TeV.  Of course, it is
an open question whether such theories can
actually be constructed, consistent with the
data.  Nevertheless, our results point to a
loophole in the common assertion that the
precision data require a Higgs boson or other
new physics to be close at hand.

The plan of this letter is as follows.  We
start by presenting the gauged chiral Lagrangian
associated with electroweak symmetry breaking.  We
then focus on the two operators that are most
important for precision measurements on the $Z$
pole.  We compute the effects of these operators
on experimental observables and derive limits
on their coefficients as a function of the scale
$\Lambda$.  Finally, we discuss our results in
the context of alternative scenarios for
electroweak symmetry breaking.

\section{Framework}

The gauged chiral Lagrangian provides a
model-independent description of the physics
that underlies electroweak symmetry
breaking.\cite{gauge,herrero}  It  is valid
for energies $E \lsim \Lambda$, where the new
physics becomes manifest.

The Lagrangian is constructed from the Goldstone
bosons $w^a$ associated with breaking $SU(2)
\times U(1) \rightarrow U(1)$.  The fields $w^a$
are assembled into the group element $\Sigma =
\exp(2i w^a \tau^a/v)$, where the $\tau^a$
are Pauli matrices, normalized to $1/2$, and $v =
256$ GeV is the scale of the symmetry breaking.
The fields $w^a$ transform nonlinearly under \two
transformations, $\Sigma\to L\Sigma R^\dagger$,
%
%
where $L \in SU(2)\equiv SU(2)_L $ and $R \in
U(1) \subset SU(2)_R$.  The gauge bosons appear
through their field strengths, $W_{\mu\nu} =
W^a_{\mu\nu} \tau^a$ and $B_{\mu\nu} = B^3_{\mu
\nu} \tau^3$, as well as through the covariant
derivative, $D_\mu \Sigma = \partial_\mu \Sigma
+ i g W^a_\mu \tau^a \Sigma - i g' \Sigma B^3_\mu
\tau^3$.

The gauged chiral Lagrangian is built from these
objects.  It can be organized in a derivative
expansion,
\be
\L = \L^{(2)} + \L^{(4)} + \ldots,
\ee
where
\bea
\L^{(2)} &=& \ \ {v^2\over 4}\, \Tr D_\mu \Sigma
D_\mu \Sigma^\dagger
   + {{g'}^2 v^2\over 16 \pi^2}\, b_1\, (\Tr
T\, \Sigma^\dagger D_\mu \Sigma )^2 \nn\\
&& \mbox{}+ {gg'\over 16 \pi^2}\,a_1\,
\Tr B_{\mu\nu} \Sigma^\dagger W_{\mu\nu} \Sigma\,,
\eea
and $T = \Sigma^\dagger \tau^3 \Sigma$.  The
Lagrangian is invariant under \two gauge
transformations.  In the unitary gauge, with
$\Sigma = 1$, the terms in $\L^{(2)}$ give
rise to the $W$ and $Z$ masses.  The terms
in $\L^{(4)}$ give rise to ``anomalous" three-
and four-gauge boson self couplings.

The coefficients $a_1$ and $b_1$ are important
because they contain information about the physics
of electroweak symmetry breaking.  Note that the
operator proportional to $a_1$ preserves weak isospin
in the limit $B^3_\mu \rightarrow 0$, while the one
proportional to $b_1$ does not.  The coefficients
are obtained by matching Green functions in the
effective theory with those of the underlying
fundamental theory, just below the scale $\Lambda$.
The coefficients $a_1$ and $b_1$ are normalized
so that they are naturally $\O(1)$ for a strongly
interacting sector with $\Lambda \simeq 3$ TeV.
They can be much smaller if the symmetry-breaking
sector is weakly coupled; they can be larger if
the fundamental theory contains many particles
charged under $SU(2)\times U(1)$.

In what follows we will study the effects of
$a_1$ and $b_1$ on the $W$ and $Z$ propagators.
These coefficients are closely related to the
parameters $S$ and $T$.  The relation is found
by renormalizing the coefficients from $\Lambda$
to the scale $M_Z$, where $S$ and $T$ are defined.
One finds
\bea
S  &=& S_0 + {1\over 6 \pi}\,\log
\left(
\Lambda\over M_Z\right) , \nn \\
T  &=& T_0 - {3\over 8 \pi c^2}\,\log
\left(
\Lambda\over M_Z\right) ,
\label{eq:ST}
\eea
where $c = \cos\theta_W$, and $S_0$ and $T_0$
are fixed in terms of $a_1$ and $b_1$ at the
scale $\Lambda$,
\be
S_0 = - {a_1\over \pi}\,, \qquad\qquad
T_0 = {b_1 \over \pi c^2}\,.
\label{eq:a1b1}
\ee
Note that the logarithms are exactly calculable
because they come from standard model loops. (We
assume explicitly that there are no light particles,
such as pseudo-Goldstone bosons, with masses between
$\Lambda$ and $M_Z$~\cite{golden}.)  Equation
(\ref{eq:ST}) connects the new physics at the scale
$\Lambda$ with precision measurements at the scale
$M_Z$.

\section{Results}

We are now ready to find the constraints imposed
by precision electroweak measurements on $S$ and
$T$, and consequently, on the scale $\Lambda$
and the coefficients $a_1$ and $b_1$.

Most global analyses of precision electroweak
data are carried out in the context of the standard
model with a fundamental Higgs boson.  Fortunately,
these analyses can be easily converted to the case
at hand.  One simply subtracts the contributions
to $S$ and $T$ from a standard model Higgs boson,
and then adds back the contribution from equation
(\ref{eq:ST}).  In this way one can readily compute
the values of $S$ and $T$ that come from the gauged
chiral Lagrangian.

The contributions to $S$ and $T$ from a heavy Higgs
have been computed in the literature.\cite{herrero}
They are
\bea
S &=& - {1\over 6 \pi}\,\Bigg[ {5\over 12}
- \log\left({M_H\over M_Z}\right)\Bigg],
\nn \\
T &=& {3\over 8 \pi c^2}\,\Bigg[ {5\over 12}
- \log\left({M_H\over M_Z}\right)\Bigg],
\eea
where the constant is computed in the \MS scheme.
Note that the logarithmic dependence on the Higgs
mass is exactly the same as the logarithmic
dependence on $\Lambda$ in equation (\ref{eq:ST}).
This is no surprise, because $M_H$ plays the role
of $\Lambda$, and the standard model renormalization
is exactly the same in each case.

With this result, we are ready to make contact
with the data.  We take
\bea
S(m_t,S_0,\Lambda)  &=&
S(m_t,M_H^{\rref};m_t^{\rref}, M_H^{\rref})
\nn\\ &&\mbox{}+ S_0 + {5 \over 72 \pi} +
{1\over 6 \pi}\,\log
\left(
\Lambda\over M_H^{\rref}\right), \nn \\
T(m_t,T_0, \Lambda) &=&
T(m_t,M_H^{\rref};m_t^{\rref},M_H^{\rref})
\label{eq:STdefb}\\
&&\mbox{}+
T_0 - {5 \over 32 \pi c^2} - {3\over 8 \pi c^2}\,
\log\left(\Lambda\over M_H^{\rref}\right) ,\nn
\eea
where $S(m_t,M_H^{\rref};m_t^{\rref}, M_H^{\rref})$
and $T(m_t,M_H^{\rref};m_t^{\rref},M_H^{\rref})$ are
the standard model values of the $S$ and $T$ parameters,
evaluated at reference values of the top quark and
Higgs boson masses, $m_t^{\rref}$, $M_H^{\rref}$.

\begin{figure}
\centerline{
\epsfxsize 3.0in
\epsfbox{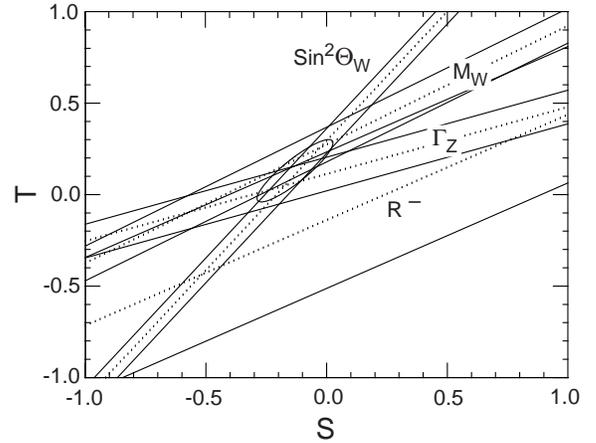}}
\vskip0.5cm
\caption{Fit to $S$ and $T$ from electroweak observables,
with $M_H^{\rref}=500\,$GeV and $m_t^{\rref}=175\,$GeV.}
\label{fig:st}
\end{figure}

We determine the physically allowed region of $S$-$T$
space from a $\chi^2$ fit to fourteen precisely measured
electroweak observables.  Each observable $\O_i$ is
represented by a four-parameter linearized function,
\bea
\O_i &=& \O_i^{\rref}+s_iS+t_iT+x_i(\alpha_s-
\alpha_s^{\rref})
\nn\\
&&\mbox{}+y_i(\Delta\alpha^5_{had}-
\Delta\alpha^5_{\rref})\,,
\eea
where $\O_i^{\rref}$ is the standard model value
of the observable at the reference values of top
quark and Higgs bosn masses.  The strong coupling
$\alpha_s$ is evaluated at the scale $M_Z$; we
take $\alpha_s^{\rref}=0.12$ as the corresponding
reference point.  In this expression, $\Delta
\alpha^5_{had}$ is the five-flavor, hadronic
portion of the vacuum polarization correction
to the electromagnetic coupling constant at the
scale $M_Z$, and $\Delta\alpha^5_{\rref}=277.5
\times10^{-4}$ is its reference point.  The
coefficients $s_i$ and $t_i$ are computed from
the standard model \cite{peskin}.  The coefficients
$x_i$, $y_i$ and the reference values $O_i^{\rref}$
are computed using the ZFITTER~6.11 computer
code~\cite{zfitter}.  All coefficients are
insensitive to the choice of the reference
points.

The fourteen observables are the width $\Gamma_Z$
of the $Z$ boson~\cite{LEPEW}; the $e^+e^-$ pole
cross section of the $Z$~\cite{LEPEW}; the ratio
of the hadronic and leptonic partial widths of the
$Z$~\cite{LEPEW}; the $Z$-pole forward-backward
asymmetries for final state leptons, $b$ quarks,
and $c$ quarks~\cite{LEPEW}; $Z$-pole left-right
coupling asymmetries for electrons and $\tau$
leptons as determined from final-state $\tau$
polarization measurements~\cite{LEPEW}; the
$Z$-pole hadronic charge asymmetry~\cite{LEPEW};
the left-right cross section asymmetry for $Z$
production~\cite{LEPEW}; the mass $M_W$ of the $W$
boson~\cite{LEPEW}; $R_-$, a quantity constructed
from the ratios of neutral- and charged-current
$\nu$ and $\bar\nu$ cross sections~\cite{nutev};
the weak charge of the Cesium nucleus~\cite{cesium};
and the weak charge of the Thallium
nucleus~\cite{thallium}.  The fit is performed
with $\Delta\alpha^5_{had}$ constrained to the
value $(277.5 \pm1.7)\times10^{-4}$, as determined
by a recent analysis~\cite{kuhn}.  The $\chi^2$
weight matrix includes correlated errors for the
LEP $Z$ lineshape parameters. The resulting
two-dimensional 68.3\% confidence region in
$S$-$T$ space is shown in Fig~\ref{fig:st} for
the reference point $(m_t^{\rref},M_H^{\rref})
= (175,500)\,$GeV.  The one-dimensional 68\%
confidence intervals for the parameters are
\begin{eqnarray}
    &&S = -0.13\pm0.10 \,, \qquad T=0.13\pm0.11
\,, \nn\\
    &&\alpha_s(M_Z) = 0.119\pm0.003\,,\nn\\
    &&\Delta\alpha^5_{had}(M_Z) =
(277.6\pm1.7)\times10^{-4}\,.
\end{eqnarray}
Note that the $S$ and $T$ confidence regions (one-
and two-dimensional) implicitly incorporate the
uncertainties resulting from the imprecise knowledge
of $\alpha_s(M_Z)$ and $\Delta\alpha^5_{had}(M_Z)$.

To test the consistency of our approach, we perform
a chi-square fit of the measured values of $S$ and
$T$ to the standard model functions $S(m_t,M_H;
m_t^{\rref},M_H^{\rref})$ and $T(m_t,M_H;m_t^{\rref},
M_H^{\rref})$, which are calculated with ZFITTER~6.11.
The $\chi^2$ weight matrix is obtained from the inverse
of the $S$-$T$ error matrix.  We add an additional
term to the $\chi^2$ function to include a constraint
on the top quark mass~\cite{tmass}, $m_t= 174.3\pm5.1$
GeV.  We then compare the result of this fit with
that of a direct fit to the standard model using the
same fourteen observables with the same constraints.
The standard model fit yields a central value for $M_H$
of 106.3~GeV and a 95\% upper limit of 228.5~GeV.  The
$S$-$T$ fit yields very consistent values of 107.4~GeV
and 228.8~GeV, respectively.

In what follows, we use a similar procedure to derive
confidence intervals for the parameters $S_0$, $T_0$,
and $\Lambda$, which characterize the alternative
electroweak symmetry breaking sector.   The measured
values of $S$ and $T$ are fit to the functions defined
in equations (\ref{eq:STdefb}),
with the same reference masses as above.  In addition,
the same $m_t$-constraining term is added to the
$\chi^2$ function.

Of course, it is not possible to determine all three
of $S_0$, $T_0$ and $\Lambda$ using just two measurements.
Indeed, for any fixed $\Lambda$, it is always possible
to adjust the matching coefficients $S_0$ and $T_0$
to fit the low energy data.  However, the situation
$S \ll S_0$ and $T\ll T_0$ would be unnatural, since
it would suggest finely tuned cancelations in equations
(\ref{eq:ST}).  Indeed, there is no reason to expect
any correlation between chiral lagrangian parameters
generated directly at the scale $\Lambda$ and logarithmic
radiative corrections generated in running the theory
from $\Lambda$ down to $M_Z$.  We will see that even
for $\Lambda\approx3\,$TeV, no such tuning is required.

The result of our fit is shown in Figure~\ref{fg:s0t0}.
We plot the allowed region for $S_0$ and $T_0$ for
$\Lambda=(3,2,1,0.5,0.1)\,$TeV.  The $\Lambda=100\,$GeV
point is shown, although our chiral Lagrangian description
is not valid for such a low cutoff.  The 68\% and 95\%
C.L.\ ellipses are shown for $\Lambda=3\,$TeV; the
fit yields the central values $(S_0,T_0)=(-0.27,0.46)$
and the 68\% C.L.\ ranges
\begin{eqnarray}
-0.37< S_0 < -0.17\,,\qquad
0.34 < T_0 < 0.58\,,
\end{eqnarray}
For smaller $\Lambda$, the central values for $S_0$
and $T_0$ become smaller, as shown in Fig.~\ref{fg:s0t0}
while the error ellipse retains its size and orientation.

\begin{figure}
\centerline{
\epsfxsize 3.0in
\epsfbox{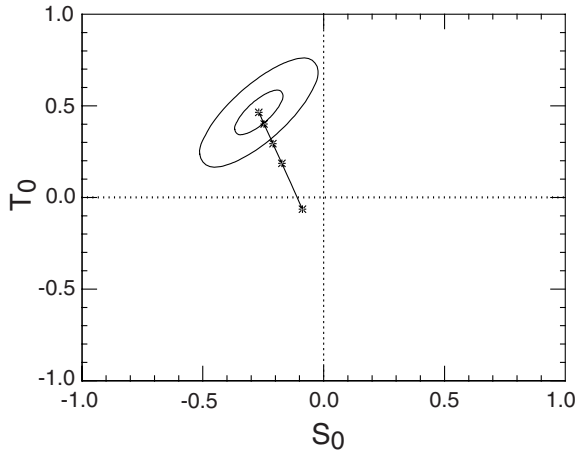}}
\vskip0.5cm
\caption{Fit to $S_0$ and $T_0$ from electroweak
observables, for $\Lambda=(3,2,1,0.5,0.1)\,$TeV. Both 68\%
and 95\% C.L.\ ellipses are shown for $\Lambda=3\,$TeV.}
\label{fg:s0t0}
\end{figure}

{}From the relation (\ref{eq:a1b1}) between $(S_0,T_0)$
and $(a_1,b_1)$, we see that chiral lagrangian coefficients
of order one or smaller are needed to fit the precision
data, for all reasonable values of $\Lambda$.  As a
measure of the tuning which is required to fit the data,
we compute the ratio of the constant term to the
logarithm in equations (\ref{eq:ST}); the deviation of
this ratio from one is an indication of the degree to
which the each constant must be adjusted to cancel the
logarithm and fit the data at $M_Z$.  Taking the central
values $(a_1,b_1)=(0.85,1.11)$ from the fit at
$\Lambda=3\,$TeV, we find a ratio of 1.4 for $S$ and
0.85 for $T$.  Even without including the experimental
uncertainties, we see that no significant tuning of $a_1$
and $b_1$ is required.

\section{Conclusion}

Precision electroweak measurements place a strong upper
limit on about 230 GeV on the mass of the Higgs boson
in the context of the standard model of particle physics.
In this letter we have seen that these measurements do
not rule out alternative theories.  Indeed, we find that
they permit strongly interacting theories with
scales as high as 3 TeV.

Nevertheless, we have seen that precision measurements
place significant constraints on these alternative
theories.  They constrain the parameters $a_1$ and
$b_1$ to be of order unity, and for $\Lambda\agt1\,$
TeV, they completely fix their signs.  It is, of
course, an urgent and open question to determine
whether a reasonable model can be constructed with
these parameters.  For example, it has previously
been observed that it is difficult to obtain $a_1>0$
in naive technicolor theories~\cite{peskin}.  In
such models, $S$ receives a small positive contribution
of approximately 0.1 for each weak doublet in the
fundamental theory.

More generally, we would argue that the data
disfavor models in which fermion masses are generated
directly by the electroweak symmetry breaking dynamics.
Fermion masses arise from interactions of the form
\be
   \Phi_U^{ij}\,\overline Q_L^iu_R^j+
   \Phi_D^{ij}\,\overline Q_L^id_R^j+
   \Phi^{ij}_L\,\overline L_L^ie_R^j\,
\ee
where $i,j=1,2,3$ are flavor indices and $\Phi_a^{ij}$,
$a=U,D,L$, are (possibly composite) fields which assume
nonzero vacuum expectation values.  In the standard model,
$\Phi_U^{ij}=\lambda_U^{ij}\Phi$,
$\Phi_D^{ij}=\lambda_D^{ij}\Phi^*$ and
$\Phi_L^{ij}=\lambda_L^{ij}\Phi$, where $\Phi$
is the single Higgs boson and the $\lambda_a^{ij}$
are 27 Yukawa couplings which break the $U(3)^5$ flavor
symmetry.  In theories in which these symmetries are
dynamically broken, the fields $\Phi_a^{ij}$ are
dyanmical degrees of freedom that carry representations
of the flavor symmetry group.  When the $\Phi_a^{ij}$
are integrated out, they give a contribution to $a_1$
which includes a trace over a large number or fields.
Generically, we expect the trace to be large:  in
the unrealistically minimal scenario in which the
trace is 27 times the contribution of a single scalar,
we find $|a_1|=27\times(5/72)=1.9$.  A more realistic
model would require significant cancelations to
achieve the observed value of $a_1$.

J.B.\ would like to thank the Aspen Center for Physics
for hospitality.  This work was supported in part by the
National Science Foundation under grants PHY--9404057 and
PHY--9604893.  Support for A.F.\ was also provided by the
NSF under National Young Investigator Award PHY--9457916,
by the DOE under Outstanding Junior Investigator Award
DE--FG02--94ER40869, by the Research Corporation under
the Cottrell Scholar program, and by the Alfred P.
\ Sloan Foundation.

{\it Note added.}  After this work was completed, we
bacame aware of Ref.~\cite{sirlin}.  In this paper
the authors claim that the upper bound on $\Lambda$
is very close to the upper bound on $M_H$ in the
standard model.  The authors of this paper neglect
$a_1$ and $b_1$, so their bound holds in the class
of models where $a_1$ and $b_1$ are near zero.

\end{document}